\documentclass[aps,preprint,tightenlines,nofootinbib]{revtex4}
\usepackage{graphicx}
\usepackage{capt-of}
\usepackage{amsmath}
\begin{document}
 \title{$^{19}\mathbf{C}$ in halo EFT: Effective-range parameters from Coulomb dissociation experiments }
\author{B.~Acharya}
\email{acharyab@phy.ohiou.edu}
\author{Daniel R. Phillips}
\email{phillips@phy.ohiou.edu}
\affiliation{Department of Physics and Astronomy and Institute of Nuclear and Particle Physics, Ohio University, Athens, OH\ 45701, USA%
\\
}

 \begin{abstract}
 We study the Coulomb dissociation of the $^{19}\mathrm{C}$ nucleus in an effective field theory that  uses the $^{18}\mathrm{C}$ core and the neutron as effective degrees of freedom and exploits the separation of scales in this halo system. We extract the effective-range parameters and the separation energy of the halo neutron from the experimental data reported in Refs.~\cite{Nakamura:1999rp,Nakamura:2003c}, taken at RIKEN by Nakamura et~al.~(1999, 2003). We obtain a value of $(575\pm55(\mathrm{stat.})\pm20(\mathrm{EFT}))$~keV for the one-neutron separation energy of $^{19}\mathrm{C}$, and $(7.75\pm0.35(\mathrm{stat.})\pm0.3(\mathrm{EFT}))$~fm for the ${}^{18}$C-neutron scattering length. The width of the  longitudinal momentum distribution predicted by EFT using this separation energy agrees well with the experimental data taken at NSCL by Bazin et al.~(1998)~\cite{Bazin:1998zz}, reaffirming the dominance of the s-wave configuration of the valence neutron. 
  \end {abstract}
  
   \maketitle
   
  \section{Introduction}
  \label{sec:intro}
 Light nuclei near the neutron drip line have generated considerable excitement in the last few decades, due, in part, to the observation of halos of neutrons in many of them \cite{Tanihata:1995yv,673808}. Such nuclei have a core of normal nuclear density surrounded by a diffuse neutron cloud that extends to distances much larger than the size of the core. Halo nuclei exhibit certain universal properties: large reaction cross section \cite{Tanihata:1986kh,Tanihata:1985zq,Ozawa:2000gx,Tanaka:2010zza}, enhanced $E1$ strength at low excitation energy \cite{Fukuda:2004ct,Kobayashi:1989vr}, and a narrow peak in the momentum distribution of the fragments in neutron removal reactions \cite{Bazin:1998zz,Tanihata:1995yv}.
Furthermore, halo nuclei in the isotopic chain of carbon are particularly interesting because  $^{20}\mathrm{C}$ and $^{22}\mathrm{C}$ have both been considered promising candidates to have bound excited Efimov states \cite{Efimov:1970zz,Mazumdar:2006tn,Canham:2008jd,Yamashita:2011cb}. In this work, we study the Coulomb dissociation of the  $^{19}\mathrm{C}$ nucleus using an effective field theory (EFT) developed in Refs.~\cite{Kaplan:1998tg,Kaplan:1998we,Gegelia:1998gn,Birse:1998dk,vanKolck:1998bw,Bertulani:2002sz,Bedaque:2003wa}, and augmented with electromagnetic interactions in Refs.~\cite{Phillips:2010dt,Rupak:2011nk,Hammer:2011ye}. This work is an application of the EFT presented in Ref.~\cite{Hammer:2011ye}, and the theory of electric dipole excitations in relativistic collisions, discussed in Refs.~\cite{Bertulani:1987tz,Bertulani:1985madeup, Bertulani:1988madeup,Bertulani:2009zk}.

 $^{18}\mathrm{C}$ has a ground state with $J^{\pi}=0^{+}$ \cite{Ajzenberg:1987}. Therefore, from the simplest shell-model picture, one would expect  $^{19}\mathrm{C}$ to have a ground state $J^{\pi}$ of $5/2^{+}$. However, based on the observation of a narrow momentum peak and a large cross section, Ref.~\cite{Bazin:1995zz} suggested that the valence neutron in $^{19}\mathrm{C}$ is in an s-wave relative to the core, consistent with a shell-model calculation based on the Warburton-Brown effective interaction \cite{Warburton:1992rh}. Ref.~\cite{Bazin:1998zz} reported that the width of the longitudinal momentum distribution of $^{18}\mathrm{C}$ produced by the Coulomb break-up of $^{19}\mathrm{C}$ did not agree with a Yukawa potential model calculation which used the $^{19}\mathrm{C}-$neutron separation energy of $160\pm110$~keV, determined from prior mass measurements. Moreover, this value was found to be incompatible with the energy spectrum  of the Coulomb dissociation cross section in Ref.~\cite{Nakamura:1999rp} for all possible configurations. The authors, therefore, revised the neutron separation energy of $^{19}\mathrm{C}$ to $530\pm130$~keV by analyzing the angular dependence of the Coulomb dissociation differential cross section. With this value, the spectrum could be well reproduced for a dominant $^{18}\mathrm{C}\left(0^{+}\right)\otimes 2s_{1/2} $ ground-state configuration with a spectroscopic factor of 0.67, leading to the assignment of $J^{\pi}=1/2^{+}$ to the $^{19}\mathrm{C}$ ground state. Ref.~\cite{Audi:2002rp}'s evaluation of $580 \pm 90$ keV for the $^{19}\mathrm{C}-$neutron separation energy is based on this result and on the value of $650 \pm 150$ keV extracted from the inclusive longitudinal momentum distribution obtained in single-neutron knockout~\cite{Maddalena:2001bn}. 
 A non-perturbative treatment of the Coulomb interaction by Banerjee and Shyam corroborated Nakamura et al.'s spin-parity and neutron separation energy assignments, albeit with a spectroscopic factor of 1 for the $^{18}\mathrm{C}\left(0^{+}\right)\otimes 2s_{1/2} $ configuration  \cite{Banerjee:1999je}. Typel and Baur studied the experiment reported in Refs.~\cite{Nakamura:1999rp,Nakamura:2003c} and concluded that higher-order electromagnetic effects are small \cite{Typel:2001mx,Typel:2008bw}. In Ref.~\cite{Singh:2008}, Singh et al. analyzed some of the data in Refs.~\cite{Nakamura:1999rp,Bazin:1998zz} using a Woods-Saxon potential between the valence neutron and the core~\footnote{We attempted to reproduce the results of Ref.~\cite{Singh:2008} in detail, but were unable to do so, even when using the same input parameters.}. A recent study hinted at the possibility that the first $5/2^+$ state of $^{19}\mathrm{C}$ is unbound \cite{Kobayashi:2011mm}.

 The first excitation energy of  $^{18}\mathrm{C}$, $1620\pm20$~keV \cite{Ajzenberg:1987}, is approximately  three times larger than the neutron separation energy of $^{19}\mathrm{C}$. When studying the excitation of ${}^{18}$C$-n$ continuum states well below this energy, the short-distance correlations inside the $^{18}\mathrm{C}$ core decouple from the long-distance part of the $^{19}\mathrm{C}$ wave function. The low-energy properties of  $^{19}\mathrm{C}$ can then be studied in an EFT which uses the $^{18}\mathrm{C}$ core and the neutron as its effective degrees of freedom. At leading order (LO) in this theory, we treat both the neutron and the core as point particles.  The effects of the finite size of the core are taken into account order by order in a systematic expansion. A rough estimate of the expansion parameter is provided by $R_{core}/R_{halo}$, the ratio of the size of the core to that of the halo. Using the values deduced in Ref.~\cite{Bazin:1995zz}, this ratio comes out to be 0.49, which is not particularly small. As a result, a LO calculation is inadequate. We perform a next-to-next-to-leading order (N$^2$LO) calculation, in which there are two undetermined parameters which need to be fitted to experimental input. 
 
This article is organized as follows. In Section~\ref{sec:theory}, we present a brief review of the relevant theory. In Section~\ref{sec:data}, we fit the tunable parameters in the EFT to data from Refs.~\cite{Nakamura:1999rp,Nakamura:2003c} to determine the values of the scattering length and the separation energy of the neutron-core system. We apply the results of Sections~\ref{sec:theory} and \ref{sec:data} in Section~\ref{sec:longmom} to obtain a prediction for the longitudinal momentum distribution, and compare it with the data of Ref.~\cite{Bazin:1998zz}. A brief conclusion is presented in Section~\ref{sec:conc}.
  
  \section{Coulomb Dissociation of $^{19}\mathbf{C}$}
  \label{sec:theory}
We present a derivation of the dipole transition strength, $B(E1)$, for the excitation of $^{19}\mathrm{C}$ to $^{18}\mathrm{C}+n$ in the continuum state, in Section~\ref{sec:eft}. This quantity is independent of the kinematics of the scattering process. A reaction theory that relates it to the Coulomb dissociation cross section is  discussed in Section~\ref{sec:reaction}.
 \subsection{$^{19}\mathbf{C}$ in Halo EFT}
  \label{sec:eft}
  We begin with a brief summary of the EFT formalism developed for $^{11}\mathrm{Be}$ in Refs.~\cite{Phillips:2010dt,Hammer:2011ye}. The same halo EFT 
with electromagnetic interactions has also been used to analyze ${}^7\mathrm{Li} + n \rightarrow {}^8\mathrm{Li} +  \gamma$~\cite{Rupak:2011nk,Fernando:2011ts} and ${}^{14}\mathrm{C} + n \rightarrow {}^{15}\mathrm{C} + \gamma$~\cite{Rupak:2012cr}. Here we specialize to the simpler case of a
neutron-core halo system in an s-wave in the absence of (shallow) p-wave bound states.
 We represent the  $^{18}\mathrm{C}$ core by a bosonic field, $c$, and the neutron by a spinor field, $n$. The $^{19}\mathrm{C}$ nucleus is constructed  to be a spinor field, $t$. With masses of the neutron, the core and the halo denoted by $m$, $M$ and $M_{nc}$, respectively, the EFT Lagrangian for this system can then be written as:
   \begin{equation}
\label{eq:lagrangian}
\begin{array}{ll}
              {\cal{L}} = n^{\dagger}\left(i \partial_{0}+\frac{\nabla^{2}}{2m}\right)n+c^{\dagger}\left(i \partial_{0}+\frac{\nabla^{2}}{2M}\right)c\\
              \qquad\qquad+\sigma t^{\dagger}\left(i \partial_{0}+\frac{\nabla^{2}}{2M_{nc}}-\Delta\right)t-g\left[t^{\dagger}nc+c^{\dagger}n^{\dagger}t\right]+...,
              \end{array}
\end{equation} where $\sigma = \pm 1$, $\Delta$ is the residual mass of the $t$-field,  $g$ is the coupling constant, and the ellipses stand for the terms which are higher order in the EFT expansion parameter. 

\vspace{1cm}

\begin{minipage}[b]{.95\textwidth}
\begin{center}
\includegraphics[scale=1.]{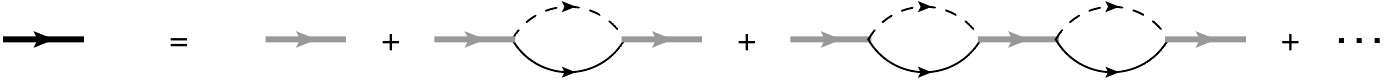}
\captionof{figure}{The expansion of the dressed $t$-propagator (thick, black). The thick gray line is the bare $t$-propagator, the dashed and the solid lines are the $c$ and the $n$ propagators respectively. The ellipses represent the rest of the terms in the infinite geometric series.\label{fig:dressedt}}
\end{center}
\end{minipage}

\vspace{1cm}

The propagator of the $t$-field is dressed by the $n$-$c$ self-energy bubbles. For a shallow s-wave state, the neutron-core binding momentum, $\gamma_0\sim k_{lo}$, where $k_{lo}$ is a generic low momentum scale of the EFT, which is of the order of  $1/R_{halo}$. Therefore all the diagrams in the right hand side of the equation depicted in Figure~\ref{fig:dressedt} need to be summed~\cite{Kaplan:1998tg,Kaplan:1998we,Gegelia:1998gn,Birse:1998dk,vanKolck:1998bw}. The dressed $t$-propagator is, therefore,
\begin{equation}
D_{t}(p)=
\dfrac{1}{{\Delta}^{(PDS)} + \sigma[p_0 - {\bf p}^2/(2 M_{nc})+i\epsilon] -\frac{\mu g^2}{2 \pi} \left[\sqrt{-2 \mu p_0+ \frac{\mu{\bf{p}}^2}{M_{nc}}-i\epsilon}-\kappa \right]},
\label{eq:dressedt}
\end{equation}
 where $\Delta^{(PDS)}$ is the residual mass of the $t$-field, renormalized in power-law divergence subtraction (PDS) scheme \cite{Kaplan:1998tg,Kaplan:1998we} with a scale $\kappa$.  Here, $\mu=mM/M_{nc}$ is the reduced mass of the neutron-core system.
 
 The transition amplitude for neutron-core scattering is given by $t(E)=g^{2}D_{t}(E,\mathbf{0})$, which we match to the second-order effective-range amplitude,
 \begin{equation}
t\left(k^{2}/2\mu\right)=\frac{2 \pi}{\mu}\frac{1}{1/a - \frac{1}{2} r_0 k^2 + i k},
\end{equation} to identify the scattering length
\begin{equation}
 a=-\left[\dfrac{2\pi \Delta^{(PDS)}}{\mu g^2}+\kappa\right]^{-1}, 
 \end{equation}
and  the effective range,
  \begin{equation}
  r_{0}=-\sigma\frac{2\pi}{\mu^{2}g^{2}}.
  \end{equation}
  The $t$ that results from Eq.~(\ref{eq:dressedt}) includes all effects up to N$^2$LO in the EFT expansion in powers of $r_0/a \sim R_{core}/R_{halo}$: the first omitted piece of the low-energy amplitude is that due to the shape parameter~\cite{Kaplan:1998tg,Kaplan:1998we,vanKolck:1998bw,Beane:2000fi}. This t-matrix then produces a bound state of binding energy $\gamma_0^2/(2 \mu)$, where 
 $\gamma_0$ is the real positive root of the equation:
\begin{equation}
\label{eq:constraint}
\frac{1}{a} + \frac{1}{2} r_{0} \gamma_{0}^2 - \gamma_{0}=0.
\end{equation}
The corresponding wave-function renormalization of the dressed $t$-propagator is
   \begin{equation}
\widetilde{Z}=\frac{2\pi\gamma_{0}}{\mu^{2}g^{2} \left(1-\gamma_{0} r_{0}\right)},
\end{equation}
which yields a radial wave function, $u_{0}(r)=\widetilde{A}\exp(-\gamma_{0}r),$ where
\begin{equation}
\label{eq:anc}
\widetilde{A}=\sqrt{\frac{2\gamma_{0}}{1-r_{0}\gamma_{0}}}
\end{equation} is the one-dimensional asymptotic normalization coefficient (ANC)~\cite{Phillips:1999hh}.

Electromagnetic interactions are now incorporated into the theory by minimal substitution: 
\begin{equation*}
\partial_{\mu}\rightarrow~D_{\mu}\equiv\partial_{\mu}+ie\hat{Q}A_{\mu},
\end{equation*} where $\hat{Q}$ is the charge operator and by adding to the Lagrangian in Equation~\eqref{eq:lagrangian}, the counter terms \cite{Hammer:2011ye}:
\begin{eqnarray}
{\cal{L}}_{ct}=-L_{E0}~t^{\dagger}\left[\nabla^{2}A_{0}-\partial_{0}(\nabla.\mathbf{A})\right]t\qquad\qquad\qquad\qquad\qquad\qquad\qquad\qquad\nonumber\\
\qquad\qquad\qquad-L_{E1}\displaystyle\sum_{j}t[(i\nabla_j n)c-i n\nabla_j c]^{\dagger}(\partial_{j}A_{0}-\partial_{0}A_{j})+\mathrm{H.C.}+...,
\label{eq:lagrangianct}
\end{eqnarray}
where H.C. stands for Hermitian conjugate. Operators involving $(\partial_{j}A_{i}-\partial_{i}A_{j})$ are not considered here because we are only interested in electric properties. 

The transition amplitude for the photodissociation of $^{19}\mathrm{C}$ into $^{18}\mathrm{C}$ and a neutron can be depicted as Feynman diagrams, as shown in Figure~\ref{fig:diagrams}. The first diagram in the series scales as $1/k_{lo}$ because the $c$ propagator scales as $1/k_{lo}^2$, and the $c\gamma c$ vertex as $k_{lo}$. Since the $nc$ vertex in the second diagram represents a p-wave interaction, it scales as  $k_{lo}^2$. The $t\gamma nc$ vertex in the third diagram scales as $k_{lo}^3$ because the $nc-$pair in the  final state is in a relative p-wave~\cite{Hammer:2011ye}. The second and the third diagrams are, therefore, suppressed by at least a factor of $R^3_{core}/R^3_{halo}$ relative to the first one. The diagrams represented by ellipses contribute at even higher orders. Hence, none of the diagrams containing $t\gamma nc$ vertices and/or final-state interactions appear at NLO or N$^2$LO. Since our s-wave amplitude is also accurate to N$^2$LO, the theoretical error of our EFT calculation is ultimately of order $R^3_{core}/R^3_{halo}$. More details on the power counting can be found in Ref.~\cite{Hammer:2011ye}.

\vspace{1cm}
\begin{minipage}[b]{.95\textwidth}
\begin{center}
\includegraphics[scale=1.]{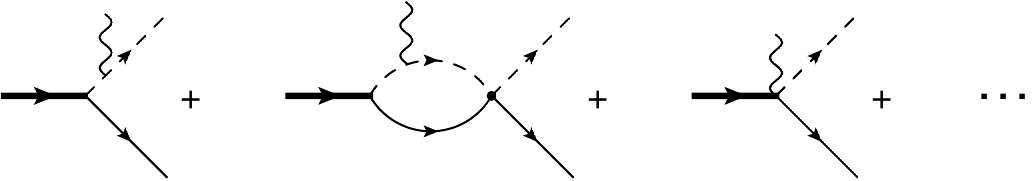}
\captionof{figure}{Diagrams contributing to the transition amplitude for the photodissociation of the halo into the core and the neutron. The ellipses represent diagrams which are higher order in the EFT expansion parameter.\label{fig:diagrams}}
\end{center}
\end{minipage}
\vspace{1cm}

With $\mathbf{q}$ the momentum of the photon and $\mathbf{p}$ the final-state momentum of the core in the CM frame of the dissociation products, the N$^2$LO amplitude can be written as 
\begin{equation}
\label{eq:amplitude}
{\cal{M}}=\sqrt{\frac{2\pi\gamma_{0}}{\mu^{2}}\frac{1}{1-r_{0}\gamma_{0}}}Ze\dfrac{1}{\frac{\gamma_{0}^{2}}{2\mu}+\frac{1}{2\mu}\left(\mathbf{p}-\frac{m}{M_{nc}}\mathbf{q}\right)^{2}},                                                                                   
\end{equation}where $Z$ is the charge of the core in the units of $e$. The binding momentum, $\gamma_{0}$, and the effective range, $r_{0}$, are both required to renormalize the dressed $t$ propagator and, therefore, need to be fixed by fitting to experimental input. 

As in Ref.~\cite{Hammer:2011ye}, we have worked in the Coulomb gauge, and used the coupling of the $c-$field to the $A_0$ photon only to derive Equation~\eqref{eq:amplitude}. $\cal{M}$ is therefore the matrix element of the plane-wave operator between the bound-state wave function and a continuum state of relative momentum $\mathbf{p}$ \cite{Hammer:2011ye}. Choosing the direction of $\mathbf{\hat{q}}$ as the $z$-axis, we obtain the matrix element of the dipole operator, $\vert \mathbf{r} \vert Y_{1}^{0}\left(\hat{r}\right)$, by picking out the term linear in {\bf q} and then dividing by $iq\sqrt{4\pi/3}$, to get
\begin{equation}
\label{eq:e1amplitude}
{\cal{M}}^{(l=1)}_{E1}=2\sqrt{\frac{6\gamma_{0}}{1-r_{0}\gamma_{0}}}\frac{m}{M_{nc}}Ze\frac{p}{\left(\gamma_{0}^{2}+p^{2}\right)^{2}}~\mathbf{\hat{p}}.\mathbf{\hat{q}}.
\end{equation}
Equation~\eqref{eq:e1amplitude} gives the amplitude in the absence of the neutron spin. We need to couple this result to the neutron spinor, and project to final states of good total angular momentum. With $\mathbf{\hat{q}}$ parallel to the z-axis, the projection of the orbital angular momentum is conserved. Also, since the $E1$ photon does not couple to the neutron, it can not cause a spin flip. The projection of the total angular momentum in the final state is therefore equal to the spin projection of the neutron in the initial state. Using properties of the Clebsch-Gordan coefficients, we then obtain
\begin{equation}
\label{eq:ms}
{\cal{M}}^{(J=3/2)}_{E1}=\sqrt{2}{\cal{M}}^{(J=1/2)}_{E1}=4\sqrt{\frac{\gamma_{0}}{1-r_{0}\gamma_{0}}}\frac{m}{M_{nc}}Ze\frac{p}{\left(\gamma_{0}^{2}+p^{2}\right)^{2}}.
\end{equation}
These matrix elements are related to $B(E1)$ by
\begin{equation}
\label{eq:be1}
\mathrm{d}B(E1)=\left(\vert{\cal{M}}^{(J=1/2)}_{E1}\vert^{2} + \vert{\cal{M}}^{(J=3/2)}_{E1}\vert^{2}\right) \frac{\mathrm{d}^{3}p}{(2\pi)^3},
\end{equation}
 which yields
 \begin{equation}
\label{eq:dbe1overdetoobig}
\frac{\mathrm{d}B(E1)}{\mathrm{d}E}= \frac{12}{\pi^{2}}\frac{\mu^{3}}{M^{2}}Z^{2}e^{2}\frac{\gamma_{0}}{1-r_{0}\gamma_{0}}\frac{p^{3}}{\left(\gamma_{0}^{2}+p^{2}\right)^{4}}.                                              
\end{equation}
Equation~\eqref{eq:dbe1overdetoobig} can also be derived by following the approach of Ref.~\cite{Rupak:2012cr}, where the authors considered the coupling of the core to the three-vector potential, $\mathbf{A}$, only  and extracted $\mathrm{d}B(E1)/\mathrm{d}E$ by calculating the photonuclear cross section for the $E1$ photon. 
 
 The expression for $\mathrm{d}B(E1)/\mathrm{d}E$ in Equation~\eqref{eq:dbe1overdetoobig} and those in Ref.~\cite{Hammer:2011ye} were obtained from matrix elements of the dipole operator calculated  with $e$ in Heaviside-Lorentz units. It is useful to express it as the ratio,
 \begin{equation}
 \label{eq:dbe1overde}
\frac{\mathrm{d}B(E1)}{e^{2}\mathrm{d}E}= \frac{12}{\pi^{2}}\frac{\mu^{3}}{M^{2}}Z^{2}\frac{\gamma_{0}}{1-r_{0}\gamma_{0}}\frac{p^{3}}{\left(\gamma_{0}^{2}+p^{2}\right)^{4}},                                                 
\end{equation} which is independent of the units of $e$ adopted. 

Our LO result, which can be obtained by setting $r_0=0$, is equivalent to that of Ref.~\cite{Bertulani:1988madeup}, where a zero-range potential model was used for the neutron-core interaction. Equation~\eqref{eq:dbe1overde} can also be recovered from the framework presented in Ref.~\cite{Typel:2001mx} (see also Ref.~\cite{Typel:2004us}), where potentials of both zero and finite range where used to assess the importance of higher-order electromagnetic effects in the break-up of $^{19}\mathrm{C}$.

  \subsection{Neutron Removal Cross Section }
  \label{sec:reaction}
We now summarize some of the results of Refs.~\cite{Bertulani:1987tz,Bertulani:1985madeup,Bertulani:1988madeup,Bertulani:2009zk} that are relevant for the case of $E1$ dominated dissociation of a s-wave one-neutron halo upon small-angle scattering by a charged target at high beam energy. In the eikonal approximation, at first order in perturbation theory, and in the absence of core-neutron final-state interactions, the dissociation cross section is given by \cite{Bertulani:1987tz}
\begin{eqnarray}
\frac{\mathrm{d}\sigma}{Q\mathrm{d}Q~\mathrm{d}^{3}p/(2\pi)^3}=24\pi^{2}\frac{Z_{t}^{2}\alpha^{2}}{\gamma^{2}\beta^{2}}\omega^{2}{Z_{eff}^{(1)}}^{2} \langle r \rangle_{01}^{2}\qquad\qquad\qquad\qquad\qquad\qquad\qquad\qquad\qquad\nonumber\\
\displaystyle \sum_{M_{1}M_{2}} i^{M_1-M_2} \chi_{M_{1}}(Q)\chi_{M_{2}}^{\ast}(Q)G_{E1M_{1}}(1/\beta)G_{E1M_{2}}^{\ast}(1/\beta)Y_{1}^{M_{1}}(\hat{p}){Y_{1}^{M_{2}}}^{\ast}(\hat{p}),
\label{eq:sigmaqp3}
\end{eqnarray}
where $Z_{t}$ is the charge of the target nucleus, $\omega$ is the excitation energy, $Q$ is the momentum exchanged between the projectile and the target nuclei, $Z_{eff}^{(1)}=mZ/M_{nc}=\mu Z/M$, $G_{E10}(x)=-i4\sqrt{\pi(x^2-1)}/3$, $G_{E11}(x)=-G_{E1-1}(x)=x\sqrt{8\pi}/3$, $\langle r \rangle_{l\lambda}= \int_{0}^{\infty} \mathrm{d}r r^{2} j_{\lambda}(pr) r R_{l}(r)$ is the radial matrix element of the operator $r$ between a bound state with radial wave function $R_{l}(r)$ and a free continuum state of orbital angular momentum $\lambda$, and
\begin{equation}
\label{eq:chi}
\chi_{M_{1}}(Q)=\int_{0}^{\infty} \mathrm{d}b~b J_{M_{1}}(Qb)K_{M_{1}}\left(\frac{\omega b}{\gamma\beta}\right)\exp(i\chi(b)),
\end{equation} where $\chi(b)$ is the eikonal phase. In the sharp-cutoff approximation,
\begin{equation}
  \exp(i\chi(b))= 
\begin{cases}
    0,& \text{if } b\leq R,\\
    1,              & \text{otherwise},
\end{cases}
\end{equation} which gives
\begin{equation}
\chi_{M_{1}}(Q)=R^{2}\int_{1}^{\infty} \mathrm{d}x~x J_{M_{1}}(QRx)K_{M_{1}}\left(\frac{\omega Rx}{\gamma\beta}\right).
\end{equation}
When expressed in terms of the reduced transition strength, $\mathrm{d}B(E1)=3 e^2 {Z_{eff}^{(1)}}^{2}  \langle r \rangle_{01}^{2}\frac{\mathrm{d}^{3}p}{(2\pi)^3}$, and integrated over $\hat{p}$ and $Q$, Equation~\eqref{eq:sigmaqp3} yields \cite{Bertulani:1987tz},
\begin{equation}
\label{eq:dsigmaoverde}
\frac{\mathrm{d}\sigma}{\mathrm{d}E}= \frac{16\pi^{3}}{9}\alpha N_{E1}(B+E,R)\frac{\mathrm{d}B(E1)}{e^{2}\mathrm{d}E},
\end{equation}where 
\begin{equation}
N_{E1}(\omega,R)= 2\frac{Z_{t}^{2}\alpha}{\pi\beta^{2}}\left(\xi K_{0}(\xi)K_{1}(\xi)-\frac{\beta^{2}}{2}\xi^{2}\left(\left(K_{1}(\xi)\right)^{2}- \left(K_{0}(\xi)\right)^{2}\right)\right),
\label{eq:photonnumb}
\end{equation} 
with $\xi=\frac{\omega R}{\gamma\beta}$ ,  is the virtual photon number for the $E1$ multipolarity, integrated over all impact parameters larger than $R$, and $B$ is the neutron separation energy.   From Equations~\eqref{eq:dbe1overde} and \eqref{eq:dsigmaoverde}, we obtain the expression for the relative energy spectrum for the Coulomb dissociation cross section at N$^2$LO in halo EFT,
\begin{equation}
\label{eq:espectrum}
\frac{\mathrm{d}\sigma}{\mathrm{d}E}= \frac{64\pi}{3}Z^{2}\alpha \frac{\mu^{3}}{M^{2}}  N_{E1}(B+E,R)\frac{\gamma_0}{1-r_0\gamma_0}\frac{p^{3}}{\left(\gamma_{0}^{2}+p^{2}\right)^{4}}.                                         
\end{equation}  
 To find the differential cross section of the center of mass of the dissociated fragments, it is convenient to replace Equation~\eqref{eq:chi} by the semiclassical approximation \cite{Bertulani:2009zk},
 \begin{equation}
 \lvert \chi_{M_{1}}(Q) \rvert^{2}=\frac{1}{Q^2}~\frac{Z_t^2Z^2\alpha^2}{4E_{cm}^2}\cot^2(\theta_{cm}/2)K_{M_1}^2\left(\xi_0\right),
 \end{equation} where $E_{cm}$ is the kinetic energy of the scattering nuclei in the center of mass frame, $\theta_{cm}$ is the Rutherford scattering angle, and $\xi_0=\omega b_0/\gamma\beta$, where
 \begin{equation}
b_0= \frac{Z_tZ\alpha}{2E_{cm}}\cot(\theta_{cm}/2)
\label{eq:ruthimpactpar}
 \end{equation}
  is the impact parameter for the Rutherford trajectory. $Q$ and $\theta_{cm}$ are related by $Q=2k_{cm}\sin(\theta_{cm}/2)$, where $k_{cm}$ is the momentum of the beam (or the target), in the projectile-target center of mass frame. Equation~\eqref{eq:sigmaqp3} then yields
 \begin{equation}
\label{eq:diffcrosssec}
\frac{\mathrm{d}\sigma}{\mathrm{d}\Omega_{cm}}= \frac{64\pi}{3}Z^{2}\alpha \frac{\mu^{3}}{M^{2}}  \frac{\gamma_0}{1-r_0\gamma_0}\int \frac{\mathrm{d}n_{E1}}{\mathrm{d}\Omega_{cm}}\frac{p^{3}}{\left(\gamma_{0}^{2}+p^{2}\right)^{4}} \mathrm{d}E,
\end{equation} where
\begin{equation}
\label{eq:diffcrosssecofphotonnumb}
\frac{\mathrm{d}n_{E1}}{\mathrm{d}\Omega_{cm}}= \frac{Z_t^2\alpha}{4\pi^2\sin^{2}(\theta_{cm}/2)}\frac{\xi_0^2}{\beta^2}\left(\frac{1}{\gamma^2}K_0^2\left(\xi_0\right)+K_{1}^2\left(\xi_0\right)\right)
\end{equation} is the number of virtual $E1$ photons per unit solid angle. Using Equation~\eqref{eq:ruthimpactpar} to relate the solid angle to the impact parameter, we can integrate Equation~\eqref{eq:diffcrosssecofphotonnumb} over all impact parameters larger than $R$ and get Equation~\eqref{eq:photonnumb}, i.e.
\begin{equation}
\int_\infty^R \frac{\mathrm{d}n_{E1}}{\mathrm{d}\Omega_{cm}}\dfrac{1}{\mathrm{d}b_0/\mathrm{d}\Omega_{cm}}\mathrm{d}b_0= N_{E1}(\omega,R).
\end{equation} 
Our use of Equations~\eqref{eq:diffcrosssec} and \eqref{eq:diffcrosssecofphotonnumb} is, therefore, equivalent to Nakamura et al.'s method of replacing $R$ in Equation~\eqref{eq:photonnumb} by $b_0$ and then differentiating it with respect to the solid angle to obtain the differential cross section.

The longitudinal momentum distribution of the dissociation cross section can be obtained from Equation~\eqref{eq:sigmaqp3} by writing the volume element $\mathrm{d}^3p$ in cylindrical coordinates with the beam direction as the $z$-axis, and integrating over the transverse coordinates. This gives
\begin{equation}
\label{eq:longmom}
 \frac{\mathrm{d}\sigma}{\mathrm{d}p_{z}} =  \frac{32\pi}{3}\alpha Z^{2}\frac{\mu^{2}}{M^{2}}   \frac{\gamma_0}{1-r_0\gamma_0}\int_{\vert p_{z}\vert}^{\infty}M_{E1}(B+E,R)\frac{p^{3}}{\left(\gamma_{0}^{2}+p^{2}\right)^{4}}\mathrm{d}p.
\end{equation}The function
\begin{eqnarray}
M_{E1}(\omega,R)=\frac{Z_{t}^{2}\alpha}{\pi\gamma^{2}\beta^{2}}\xi^{2}\left(\left(K_{1}(\xi)\right)^{2}- \left(K_{0}(\xi)\right)^{2}\right) \left(1+2P_{2}(p_{z}/p) -\gamma^{2}\left(1-P_{2}(p_{z}/p)\right) \right) \nonumber \\
+\frac{2Z_{t}^{2}\alpha}{\pi\beta^{2}}\xi K_{1}(\xi)K_{0}(\xi) \left(1-P_{2}(p_{z}/p)\right)
\end{eqnarray} contains all required information about the kinematics of the scattering.

  \section{Extracting Parameters from Coulomb Dissociation Data}
  \label{sec:data}
  There are two parameters in Equations~\eqref{eq:espectrum} and \eqref{eq:diffcrosssec}, $r_0$ and $\gamma_0$, which cannot be calculated within the theory. At this order in EFT, the $^{18}$C-$n$ scattering length, $a$, is constrained by Equation~\eqref{eq:constraint}, and the ANC, $\widetilde{A}$, by Equation~\eqref{eq:anc}.  We can thus pick any two of the four parameters $\gamma_0$, $a$, $r_0$, and $\widetilde{A}$
as the undetermined ones and fit those to experimental input. The neutron separation energy, $B$, is then equal to $\gamma_0^2/2\mu$. 
  
 \subsection{Angular Distribution}
 
 The Coulomb dissociation of $^{19}\mathrm{C}$ into $^{18}\mathrm{C}~+~n$ on a $^{208}\mathrm{Pb}$  target at $67~A$~MeV was studied in Ref.~\cite{Nakamura:1999rp}. With wave functions calculated from a 
 Woods-Saxon potential, the neutron separation energy was determined to be  $530\pm130$~keV from an analysis of the angular distribution of the $^{18}\mathrm{C}~+~n$ center of mass. By fitting the differential cross section given by Equation~\eqref{eq:diffcrosssec} to the same data, we determine not only the neutron separation energy, but also the  $^{18}{\text C}$-$n$ scattering length.
 
  We convolve the differential cross section in Equation~\eqref{eq:diffcrosssec} with the angular resolution of the detector and fit  $a$ and $B$ to the data depicted in Figure 2 of Ref.~\cite{Nakamura:1999rp}. To minimize the contribution of nuclear interaction to the break-up, we exclude the small impact parameter ($\theta>2.2$~deg.) data points. A minimum $\chi^{2}$ of 1.23 per degree of freedom is obtained at  $B$=540~keV and $a$=7.5~fm. In Figure~\ref{fig:angularcontour}, we show the contour plot for $\Delta\chi^{2}$=1 in the $aB$-plane. The projections of the contour on the coordinate axes give the 1-$\sigma$ confidence intervals on the marginalized probability distributions of $a$ and $B$. The 1-$\sigma$ interval for the separation energy is found to be $(480,630)$~keV and that for the scattering length is $(6.9,8.0)$~fm. This is consistent with the value $B=530\pm130$~keV, determined by Nakamura~et~al. from the same data.

\vspace{1cm}

\noindent\begin{minipage}[b]{0.49\textwidth}
\begin{center}
\includegraphics[width=0.8\textwidth]{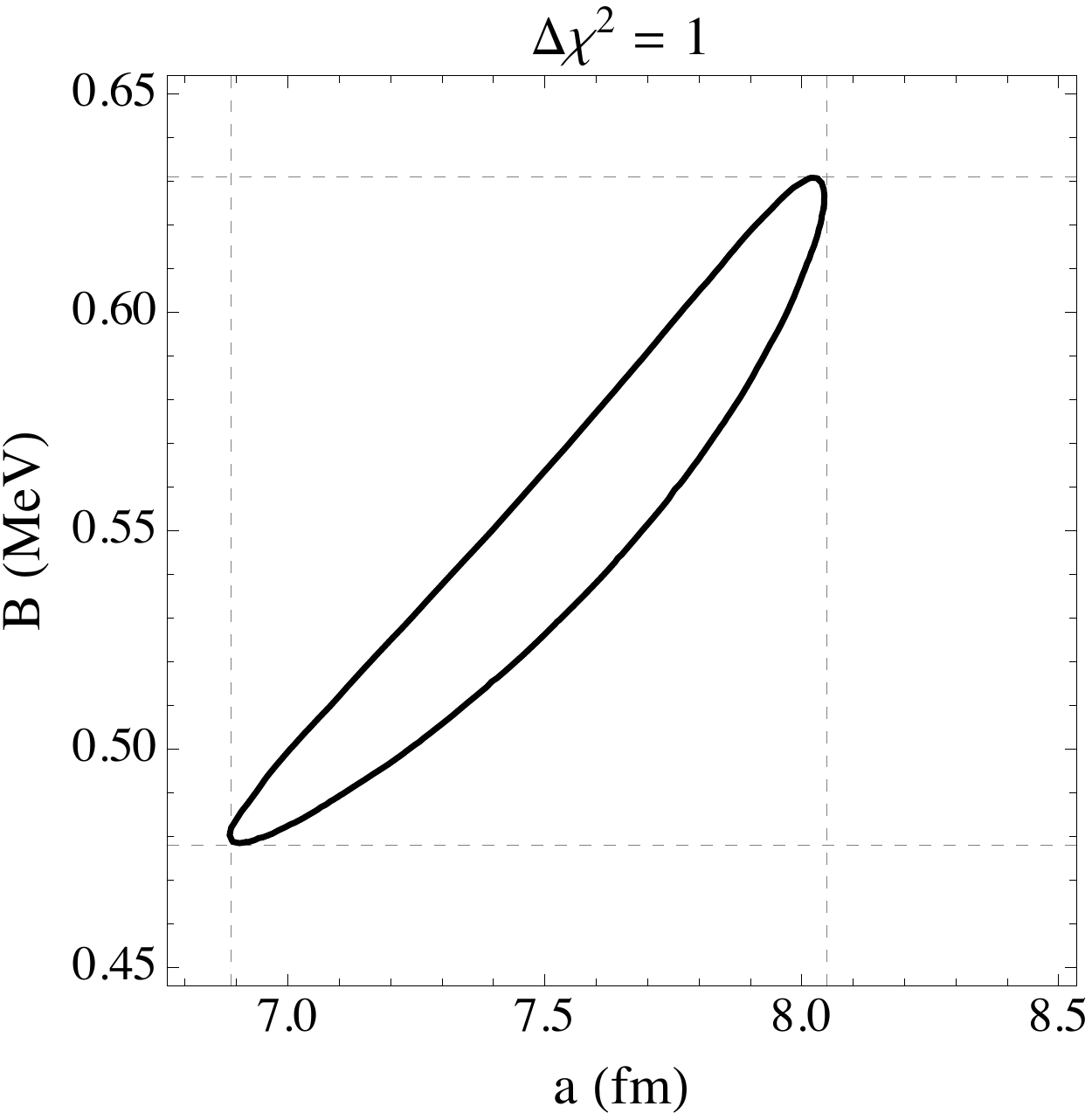}
\captionof{figure}{Contour plot of $\Delta\chi^{2}$=1 for the angular distribution of the differential cross section in the $aB$-plane.}
\label{fig:angularcontour}
\end{center}
\end{minipage}~
\begin{minipage}[b]{0.49\textwidth}
\begin{center}
\includegraphics[width=0.8\textwidth]{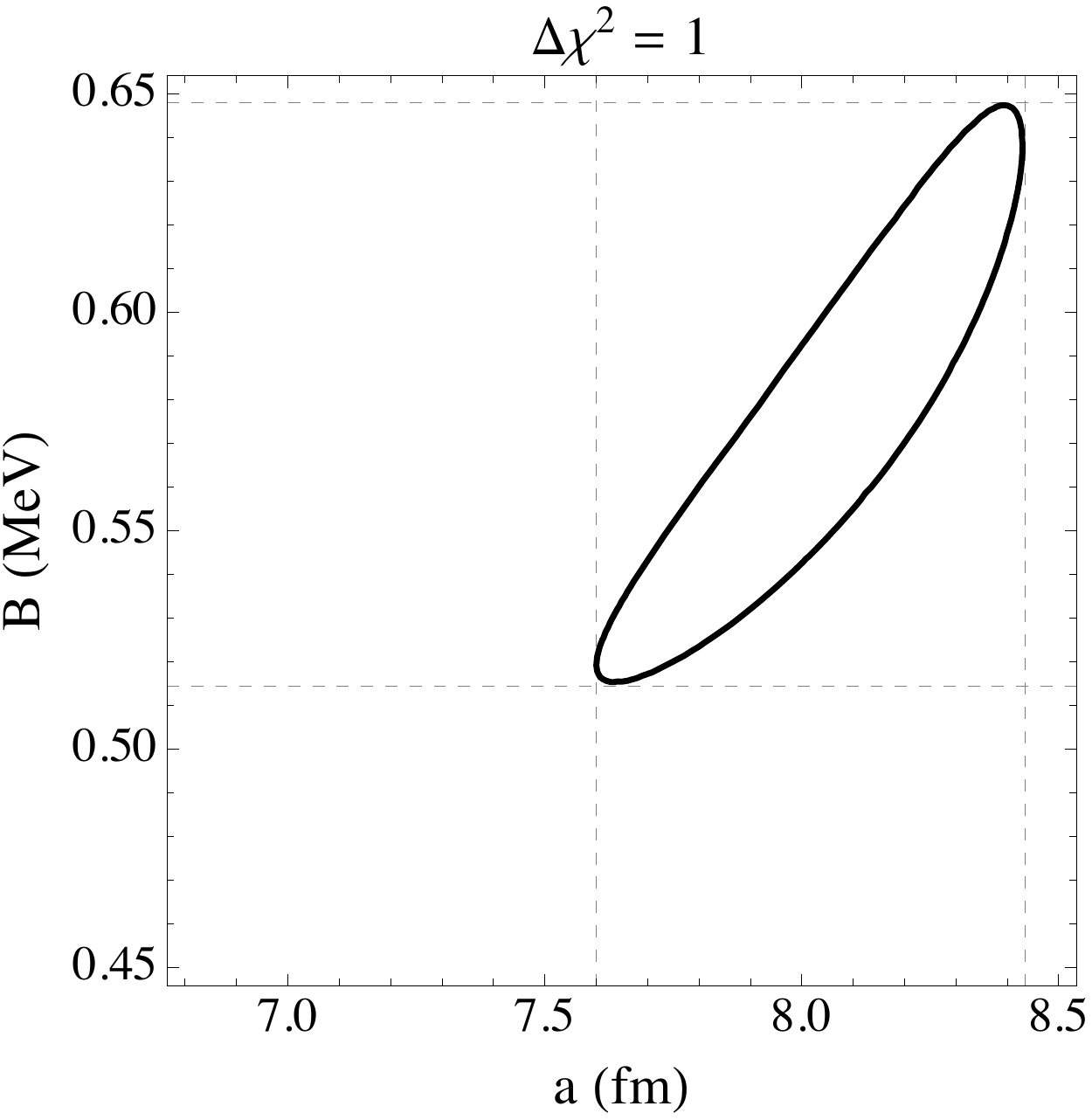}
\captionof{figure}{Contour plot of $\Delta\chi^{2}$=1 for the energy spectrum of the total cross section in the $aB$-plane.}
\label{fig:spectrumcontour}
\end{center}
 \end{minipage}
 
 \vspace{1cm}
  
   \subsection{Relative Energy Spectrum} 

Citing the sensitivity of Equation~\eqref{eq:dsigmaoverde} to the final-state interactions of $^{18}\mathrm{C}$ and $n$, Nakamura~et al. did not use the energy spectrum of the total cross section in the determination of the neutron separation energy in Ref.~\cite{Nakamura:1999rp}. However, we have shown in Section~\ref{sec:eft} that diagrams involving final-state interactions do not appear up to N$^2$LO in our theory.
We, therefore, proceed to fit the energy spectrum of the cross section given by Equation~\eqref{eq:espectrum} to the data in Figure 2(b) of Ref.~\cite{Nakamura:2003c}. Note that this data was obtained in  the same experiment reported in Ref.~\cite{Nakamura:1999rp}, but with a different method to deal with the nuclear contribution to the break-up process: Ref.~\cite{Nakamura:1999rp} subtracted a background scaled from $^{12}\mathrm{C}$-target spectrum whereas Ref.~\cite{Nakamura:2003c} used a large value of the impact-parameter cut ($R$=30~fm). 

 We convolve Equation~\eqref{eq:espectrum} with the spectral resolution of the detector and fit $a$ and $B$ to the data. We exclude the $E>1~\mathrm{MeV}$ data points because these lie outside the domain of convergence of the EFT. A minimum $\chi^{2}$ of 1.7 per degree of freedom is obtained at  $B$=580~keV and $a$=8.1~fm. The 1-$\sigma$ intervals for $B$ and $a$ from Figure~\ref{fig:spectrumcontour}, $(510,650)$~keV and $(7.6,8.4)$~fm, respectively,  are consistent with the those determined from the angular distribution.

The data in Refs.~\cite{Nakamura:1999rp}~and~\cite{Nakamura:2003c}  were obtained from the same experiment. However, we render the data sets independent by removing an overlapping region form our analysis. We can, then, simply add the $\chi^{2}$'s. The combined data has a minimum $\chi^{2}$ of 1.27 per degree of freedom at $B$=575~keV and $a$=7.75~fm. These correspond to a ${}^{18}$C-$n$ effective range of 2.6~fm. From Figures~\ref{fig:combinedcontour} and \ref{fig:contourar}, the 1-$\sigma$ confidence intervals for $a$, $B$ and $r_0$ are determined to be $(7.4,8.1)$~fm, $(520,630)$~keV, and $(1.7,3.2)$~fm, respectively.

 The ratio $r_0/a=0.33$ of the best fit to the combined data set provides a more accurate value for the EFT expansion parameter than our initial estimate $R_{core}/R_{halo}\sim0.49$. Therefore, in addition to the statistical errors, all the parameters determined above have a relative error of $r_0^3/a^3=0.036$.  The EFT extraction thus gives values of $B$, $a$, and $r_0$ of $(575\pm55(\mathrm{stat.})\pm20(\mathrm{EFT}))$~keV, $(7.75\pm0.35(\mathrm{stat.})\pm0.3(\mathrm{EFT}))$~fm, and $(2.6^{+0.6}_{-0.9}(\mathrm{stat.})\pm0.1(\mathrm{EFT}))$~fm, respectively. Note that the second errors quoted represent the uncertainty in the EFT calculation. Uncertainties in the reaction theory discussed in Sec.~\ref{sec:reaction} must be assessed separately.
 \vspace{1cm}

 \noindent\begin{minipage}[b]{0.49\textwidth}
\begin{center}
\includegraphics[width=0.8\textwidth]{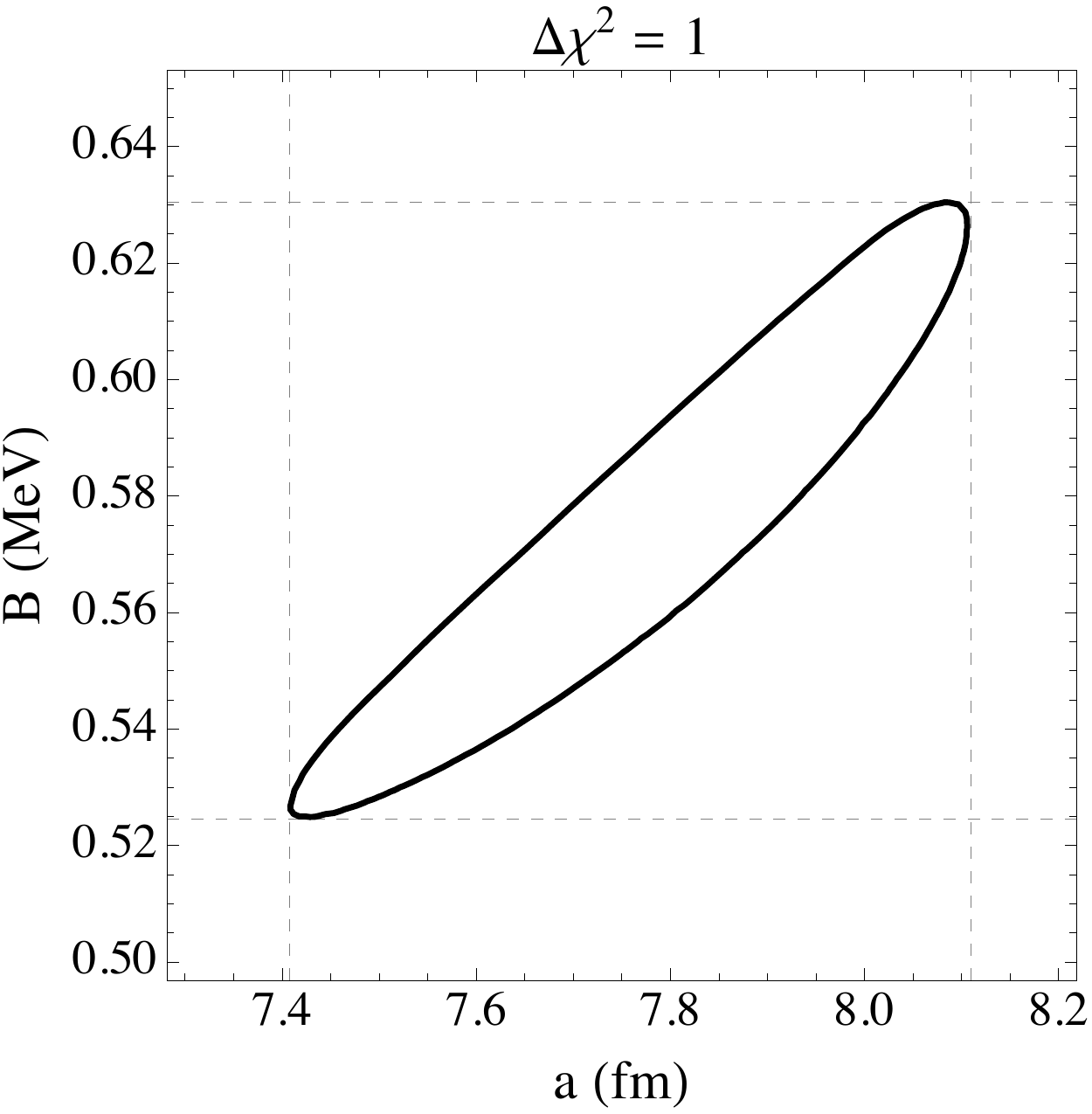}
\captionof{figure}{Contour plot of $\Delta\chi^{2}$=1 for the combined data in the $aB$-plane.}
\label{fig:combinedcontour}
\end{center}
\end{minipage}~
\begin{minipage}[b]{0.49\textwidth}
\begin{center}
\includegraphics[width=0.8\textwidth]{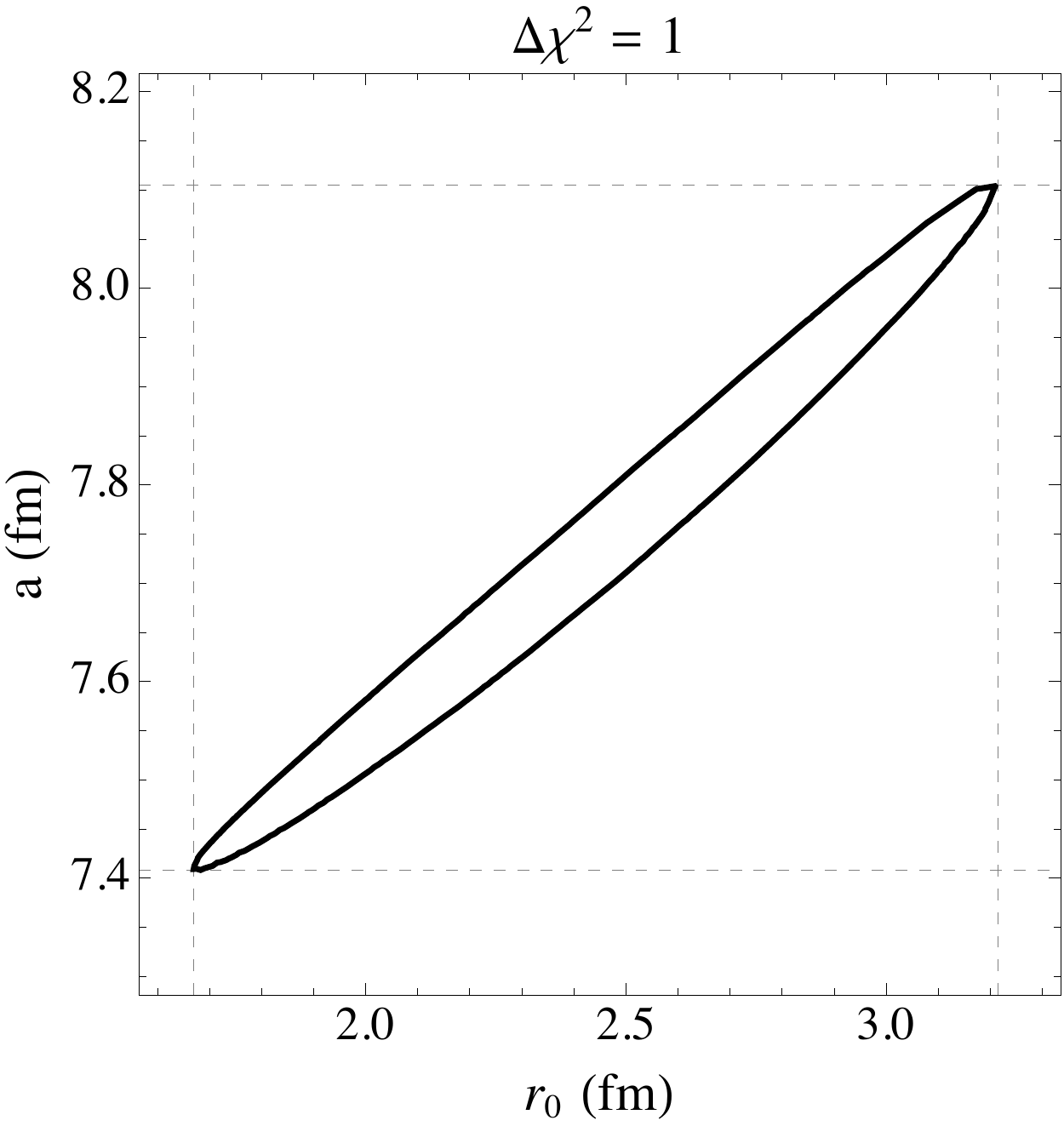}
\captionof{figure}{Contour plot of $\Delta\chi^{2}$=1 for the combined data in the $ar_0$-plane.}
\label{fig:contourar}
\end{center}
 \end{minipage}
 
 \vspace{1cm}
 
In Figures~\ref{fig:angular} and \ref{fig:spectrum}, we show the input data along with the best fits. The dashed line in Figure~\ref{fig:angular} (Figure~\ref{fig:spectrum}) is the fit to the small-angle (low-energy) data shown in the same figure, and the solid line is the fit to the combined data set. The agreement is very good, and in the case of Figure~\ref{fig:spectrum} it extends even beyond the fit region $E < 1$ MeV.

\vspace{1cm}

 \noindent\begin{minipage}[b]{0.49\textwidth}
\begin{center}
\includegraphics[width=0.8\textwidth]{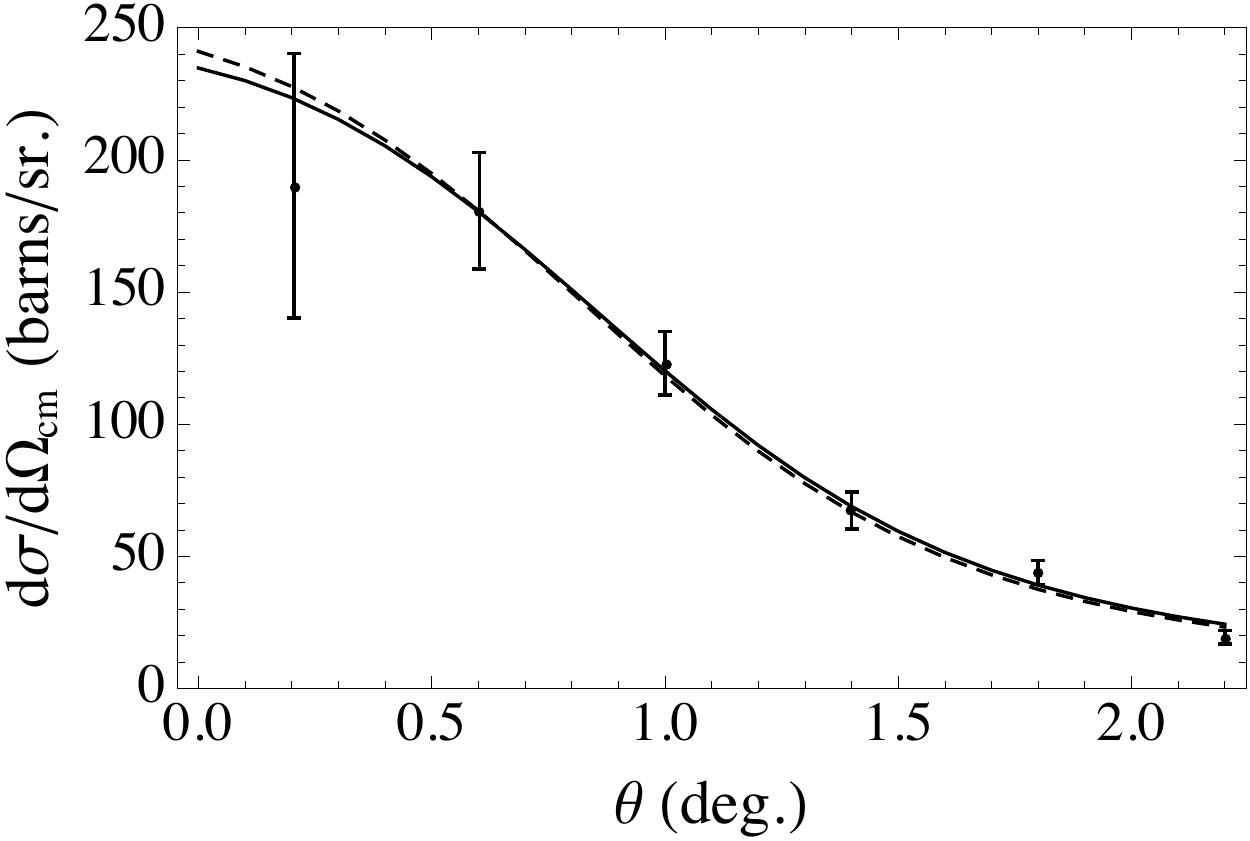}
\captionof{figure}{Angular distribution of the differential cross section at \emph{a}=7.5 fm and \emph{B}=540 keV (dashed), and at \emph{a}=7.75 fm and \emph{B}=575 keV (solid). Data from Ref.~\cite{Nakamura:1999rp}.}
\label{fig:angular}
\end{center}
\end{minipage}~
\begin{minipage}[b]{0.49\textwidth}
\begin{center}
\includegraphics[width=0.8\textwidth]{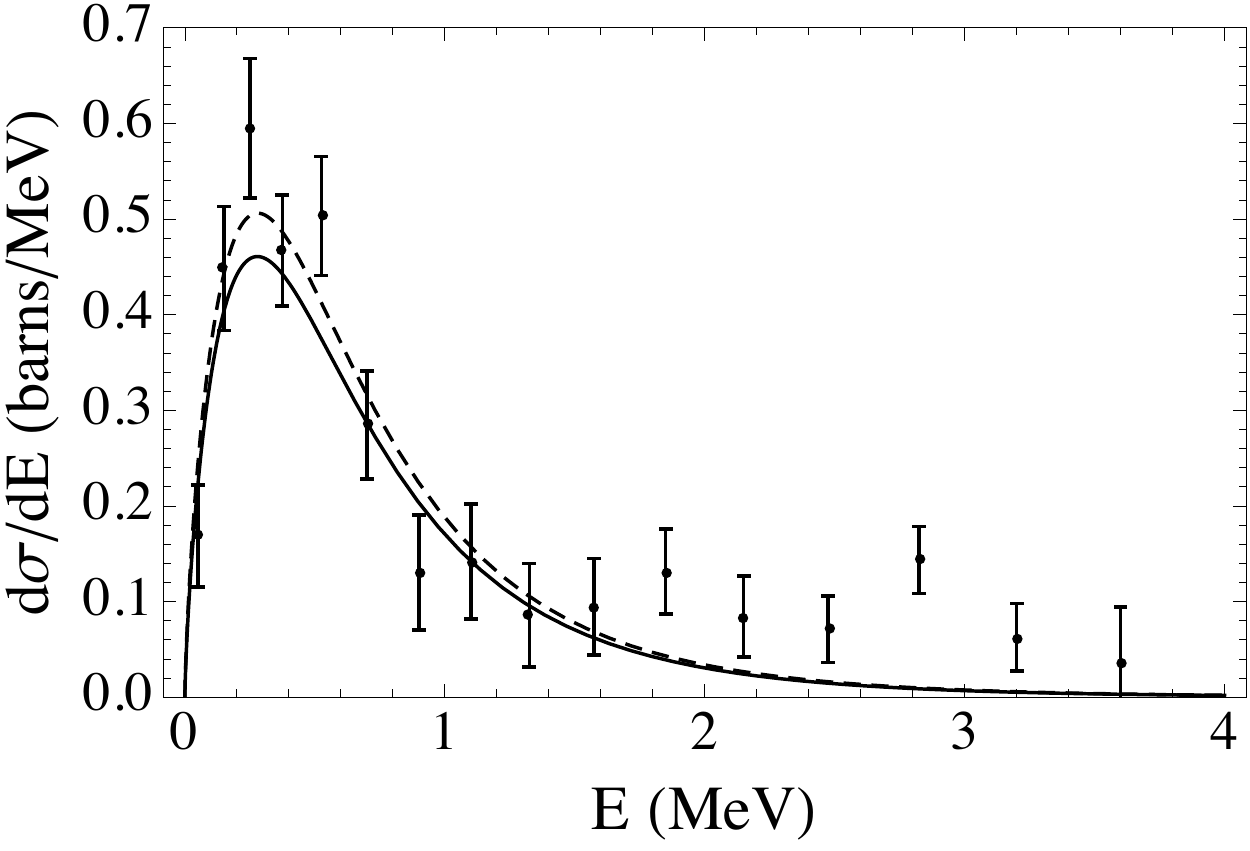}
\captionof{figure}{Relative energy spectrum of the differential cross section at \emph{a}=8.1 fm and \emph{B}=580 keV (dashed), and at \emph{a}=7.75 fm and \emph{B}=575 keV (solid). Data from Ref.~\cite{Nakamura:2003c}.}
\label{fig:spectrum}
\end{center}
 \end{minipage}

\vspace{1cm} 
 
  \section{Longitudinal Momentum Distribution}
  \label{sec:longmom} 
 In the sudden approximation \cite{Serber:1947zz}, the momentum distribution of the fragments of a break-up reaction directly reflects the bound state wave function. The presence of a narrow peak in the momentum distribution of the products of neutron removal reactions has, therefore, been widely used to establish the halo nature of nuclei \cite{Tanihata:1995yv,Bazin:1995zz,Marques:1996,Baumann:1998pm,Fang:2004et,Ozawa:2011zz}. In Ref~\cite{Bazin:1998zz}, the longitudinal momentum distribution of $^{18}\mathrm{C}$ after Coulomb dissociation of $^{19}\mathrm{C}$ on a $^{181}\mathrm{Ta}$ target was studied at a beam energy of 88~MeV/u (see also the earlier Ref.~\cite{Marques:1996}).  The data can be compared to the prediction of Equation~\eqref{eq:longmom}, with the value of $B$ determined in Section~\ref{sec:data}, up to an overall normalization factor (because the experimental result is given in arbitrary units). Since the data is given as a function of the fragment momentum measured in the laboratory frame of reference, we need to apply a Lorentz boost to $p_z$. However this gives us a peak position which is different from that seen experimentally by about 2\%, which appears to be consistent with the experimental uncertainty in the absolute energy calibration and particle energy loss in the target foil. We, therefore, fit the position and the height of the predicted peak to the data. (Note that this is different to what was done in Ref.~\cite{Singh:2008}, where the authors aligned their peak with the highest of the experimental data points.) 
 
 In Figure~\ref{fig:longmom}, we show the longitudinal momentum distribution of the Coulomb break-up cross section given by Equation~\eqref{eq:longmom} at $B$=575~keV and $r_{0}$=2.6~fm for a $^{181}\mathrm{Ta}$ target. We used $R$=13~fm, obtained by adding the nuclear radii of the projectile and the target, and a small correction to account for the bending of the Rutherford trajectory. The width of the curve predicted by Equation~\eqref{eq:longmom}, with a separation energy of 575~keV, shows good agreement with the data in Figure~\ref{fig:longmomdata}. This reinforces the notion that the low-momentum  part of the $^{19}\mathrm{C}$ wave function is dominated by a configuration with a  loosely bound neutron and a $^{18}\mathrm{C}$ core in a relative s-wave. 

 \noindent\begin{minipage}[b]{0.49\textwidth}
\begin{center}
\includegraphics[width=0.8\textwidth]{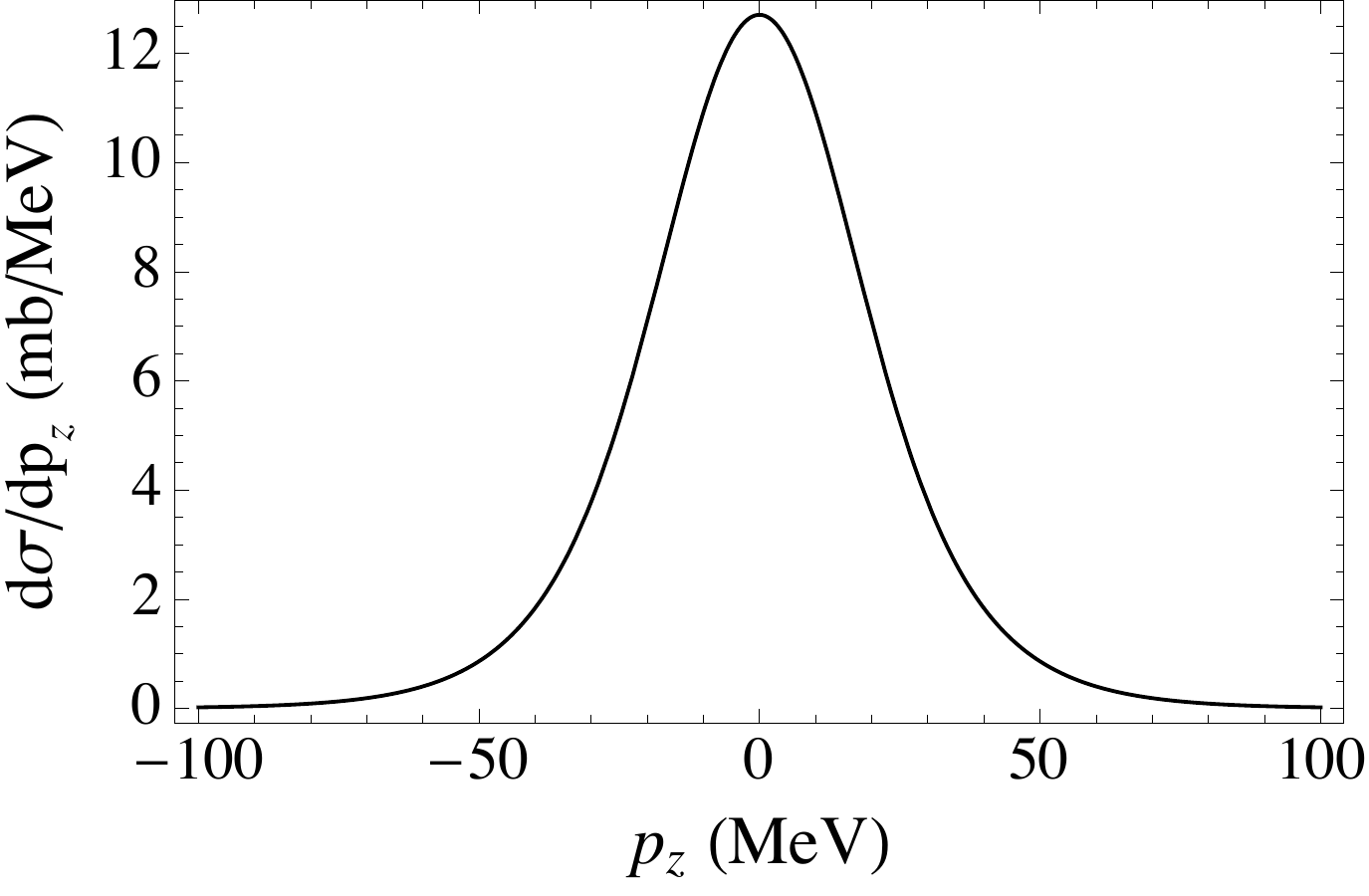}
\captionof{figure}{Longitudinal momentum distribution of the dissociation cross section on a  $^{181}\mathrm{Ta}$ target at 88~MeV/u for \emph{B}=575~keV and $r_0$=2.6~fm with $R$=13~fm.}
\label{fig:longmom}
\end{center}
\end{minipage}~
\begin{minipage}[b]{0.49\textwidth}
\begin{center}
\includegraphics[width=0.8\textwidth]{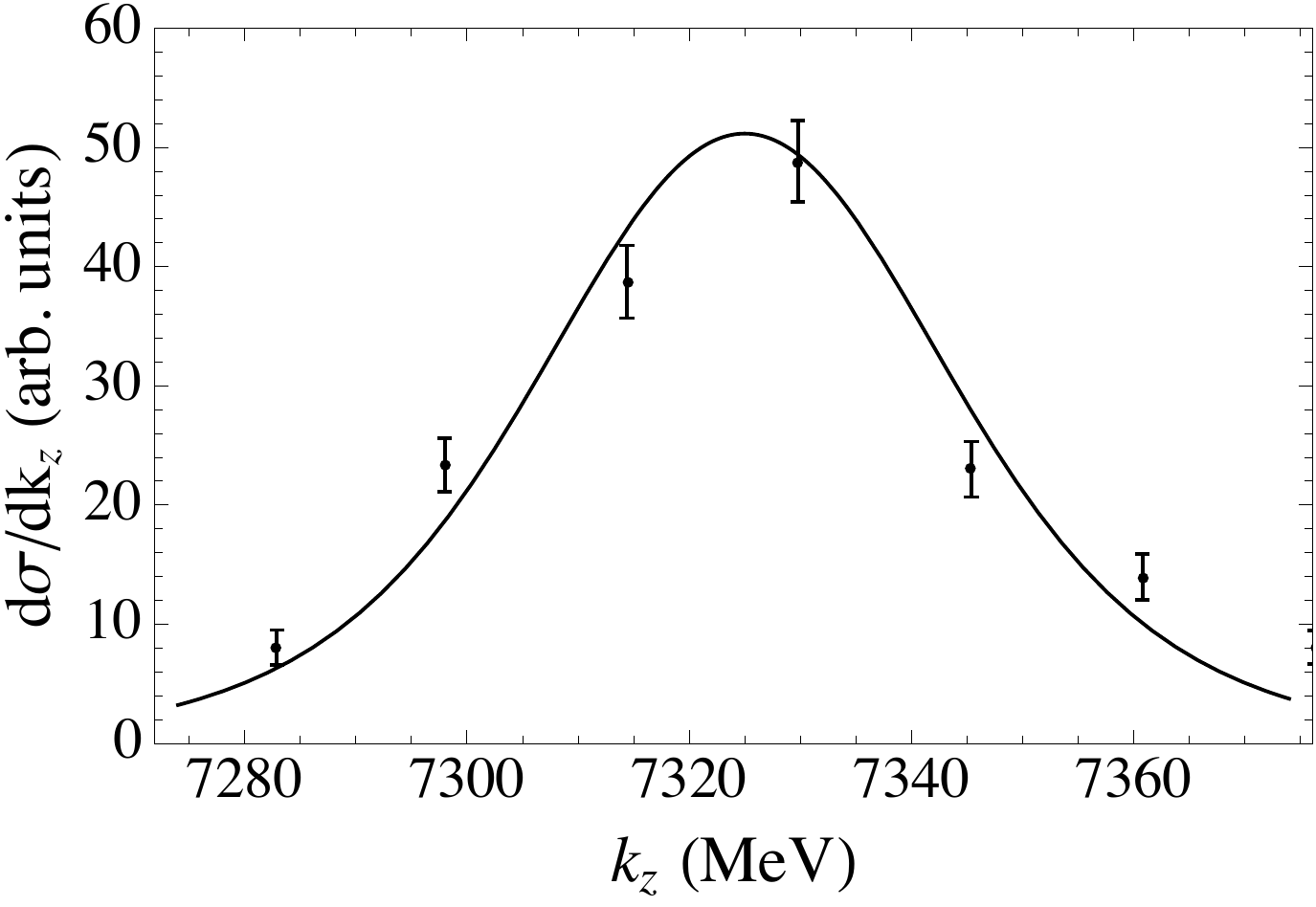}
\captionof{figure}{Longitudinal momentum distribution of the dissociation cross section with the normalization and the peak position fitted to data from Ref.~\cite{Bazin:1998zz}. }
\label{fig:longmomdata}
\end{center}
 \end{minipage}
  
  \section{Conclusion}
  \label{sec:conc}
  
An EFT analysis of data on the Coulomb dissociation of $^{19}\mathrm{C}$ from Refs.~\cite{Nakamura:1999rp,Nakamura:2003c} gives a neutron separation energy of $(575\pm55(\mathrm{stat.})\pm 20(\mathrm{EFT}))$~keV, in agreement with the previously determined values of $530\pm130$~keV \cite{Nakamura:1999rp} and $580\pm90$~keV  \cite{Audi:2002rp}. The width of the longitudinal momentum distribution obtained in Coulomb dissociation of ${}^{19}$C is then predicted by EFT and shows  good agreement with data.

The quantities determining these experimental observables are the $^{19}\mathrm{C}-$neutron separation energy and the asymptotic normalization coefficient of the associated ${}^{18}$C$-n$ wave function. We do not compute, nor do we need to, the spectroscopic factor of the s-wave ${}^{18}$C-$n$ configuration in the ${}^{19}$C ground state. At the EFT order to which we work this component of the ${}^{19}$C wave function is the only one that enters the amplitude for Coulomb dissociation at low energies. 

Furthermore, since the neutron separation energy and ANC constrain the effective-range parameters of ${}^{18}$C-$n$ scattering we can infer that the scattering length of the neutron and core is $(7.75\pm0.35(\mathrm{stat.})\pm0.3(\mathrm{EFT}))$~fm and the  $^{18}\mathrm{C}-n$ effective range is $(2.6^{+0.6}_{-0.9}(\mathrm{stat.})\pm0.1(\mathrm{EFT}))$~fm. The theoretical uncertainty stemming from the reaction theory used to connect the E1 matrix element to the observed cross sections is not included in the error bars quoted here. Corrections due to higher-order electromagnetic effects between the ${}^{208}$Pb target and the ${}^{19}$C beam is a subject for further investigation. However, these corrections are expected to be small \cite{Typel:2001mx,Typel:2008bw}.

Since the expansion parameter inferred from the extracted values of the effective range and scattering length is $\approx 0.3$ the EFT's convergence is quite good in this halo system. Our N$^2$LO calculation has uncertainties that are markedly better than the statistical precision of the data. The expansion parameter is a little smaller than the nominal expansion parameter $R_{core}/R_{halo} \approx 0.49$ obtained from the size of ${}^{18}$C.
  
The break-up of $^{19}\mathrm{C}$ has also been studied in experiments with light targets \cite{Maddalena:2001bn,Baumann:1998pm,Chiba:2004madeup, Kobayashi:2011mm}. In these studies the height of the peak of the momentum distribution was found to be of the order of a millibarn/MeV. Our result for the peak height suggests that Coulomb break-up accounts for roughly 10\% of this cross section. Similar studies using a target with a higher charge could facilitate comparison.
  
  \acknowledgments{This work was supported by the US~Department of Energy under grant DE-FG02-93ER40756. 
  We thank T.~Nakamura for sending us the data from Refs.~\cite{Nakamura:1999rp} and \cite{Nakamura:2003c}, for explaining details of the experiment to us, and for comments on the manuscript. We are grateful to G.~Baur for useful correspondence regarding Refs.~\cite{Typel:2001mx} and \cite{Typel:2008bw}, to D.~Bazin and R.~Kanungo for discussions regarding data, to R.~Kharab for providing details of the calculations reported in Ref.~\cite{Singh:2008}, and to H.-W. Hammer for suggesting $^{19}$C as a good candidate for the application of halo EFT.}
   
\end{document}